\documentclass[10pt, conference]{IEEEtran}
\usepackage[utf8]{inputenc}
\usepackage{algorithm,algcompatible,amssymb,amsmath}

\algnewcommand\algorithmicto{\textbf{to}}
\algnewcommand\RETURN{\State \textbf{return} }
\usepackage[margin = 0.7in]{geometry}
\usepackage[utf8]{inputenc}

\usepackage{amsthm}
\usepackage{amsmath}
\usepackage{amssymb}
\usepackage{amsfonts}
\usepackage{amscd}
\usepackage{mdframed}
\usepackage{mathrsfs}
\usepackage{graphicx}
\usepackage{bm}
\usepackage{url}
\usepackage{algorithm}
\usepackage{verbatim}
\usepackage{algpseudocode}
\usepackage[mathcal]{euscript}
\usepackage{booktabs}
\usepackage[normalem]{ulem}
\usepackage{caption}
\usepackage{subcaption}
\usepackage{float}
\usepackage{tabularx}
\usepackage{multicol}
\usepackage{graphicx}
\algnewcommand{\LineComment}[1]{\State \(\#\) #1}

\usepackage{mathtools}

\DeclarePairedDelimiter\floor{\lfloor}{\rfloor}

\newtheorem{theorem}{Theorem}
\newtheorem{lemma}{Lemma}

\newcommand{\nn}{\nonumber}

\begin{document}
	
\title{Self-Sustainable Key Generation: Strategies and Performance Bounds under DoS Attacks}
	
\author{Rusni Kima Mangang and J. Harshan\\
Department of Electrical Engineering,\\ 
Indian Institute of Technology Delhi, India.
}
		
\maketitle
	
\begin{abstract}
Denial-of-Service (DoS) threats pose a major challenge to the idea of physical-layer key generation as the underlying wireless channels for key extraction are usually public. Identifying this vulnerability, we study the effect of DoS threats on relay-assisted key generation, and show that a reactive jamming attack on the distribution phase of relay-assisted key generation can forbid the nodes from extracting secret keys. To circumvent this problem, we propose a self-sustainable key generation model, wherein a frequency-hopping based distribution phase is employed to evade the jamming attack even though the participating nodes do not share prior credentials. A salient feature of the idea is to carve out a few bits from the key generation phase and subsequently use them to pick a frequency band at random for the broadcast phase. Interesting resource-allocation problems are formulated on how to extract maximum number of secret bits while also being able to evade the jamming attack with high probability. Tractable low-complexity solutions are also provided to the resource-allocation problems, along with insights on the feasibility of their implementation in practice. 


\begin{IEEEkeywords}
Key generation, denial of service threats, jamming, countermeasures
\end{IEEEkeywords}

\end{abstract}
\section{Introduction}

While 6G networks are expected to deliver massive, seamless and secure connectivity through intelligent wireless devices, their deployment is also expected to open new attack surfaces for adversarial entities. If such attack surfaces are not carefully handled in the ongoing research activities towards their protocol stack, it may not deliver the promised security features. Furthermore, given that physical-layer security solutions are expected to make inroads into 6G, it is timely to identify new attack surfaces on their key management and authentication protocols, and provide appropriate solutions. 

With regards to key management, physical-layer secret key generation is a promising technique to ensure confidentiality and integrity for communication between two devices in 6G \cite{mahg}. The unique feature of secret key generation stems from the inherent randomness and reciprocity of the wireless channels, along with providing uncorrelated channel characteristics at an eavesdropper. Despite its effectiveness in harvesting secret-keys, it is worth mentioning that the frequency bands used for probing and reconciliation are usually public. This is because the wireless nodes do not have any shared common randomness to start with. As a consequence, the above scheme is vulnerable to Denial-of-Service (DoS) attacks during the phases of probing and reconciliation of key generation. As a straightforward attack, injecting jamming energy during the probing phase can disrupt the reciprocity in the effective wireless channel, thereby forbidding the nodes to generate a secret-key. For some recent DoS attacks on key generation, we refer the readers to \cite{tmag}-\cite{mitev}. While the problem of designing key generation protocols under DoS attacks is at this nascent stages, similar questions on designing key generation protocols in settings such as multi-party group secret key generation or relay-assisted key generation are still unsolved. In this work, we address the vulnerability of relay-assisted key generation methods to DoS threats, and answer a variety of questions on designing \emph{self-sustainable} countermeasures that continue to synthesize secret keys by evading the DoS threat. 

\begin{figure}
	\begin{center}
		\includegraphics[width = \linewidth]{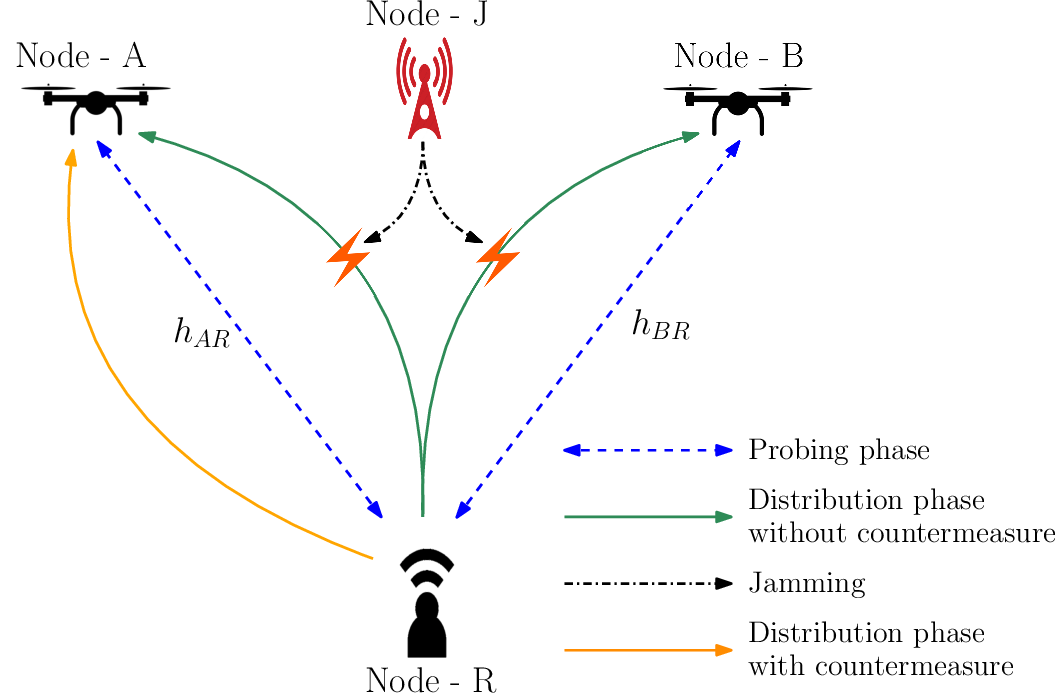}
	\end{center}
	\vspace{-0.1cm}
    \linespread{1}\selectfont{}
	\caption{Network model comprising Node-A and Node-B which seek the help of Node-R to harvest secret keys. We propose a novel countermeasure when the key distribution phase of this network is jammed by the active adversary Node-J.}
    \label{fig:network model} 
\end{figure}

We consider a relay-assisted key generation model, wherein two legitimate nodes, denoted by Node-A and Node-B, that have no direct channel between them (or insufficient randomness in their wireless channel), take the help of a trusted relay, denoted by Node-R, to generate secret keys. In a typical setting without the presence of active adversaries, secret keys are generated between Node-A and Node-B as follows. Each node generates a secret-key with Node-R during the so-called key generation phase. After this follows the distribution phase, wherein Node-R distributes an XOR version of the generated keys to both the nodes. This way, Node-A can extract the key generated between Node-B and Node-R, and then use it to secure its communication with Node-B. In the context of this work, while a DoS threat can be launched on any phase of the protocol, we are interested in studying the consequences of a jamming attack on the distribution phase. In such a case, Node-A cannot recover the XOR version broadcast by Node-R, and therefore, it will not be able to synthesize a secret key with Node-B. With that, the problem statement of interest
is \emph{How to assist key generation between Node-A and Node-B
when the broadcast phase of the protocol is subject to jamming
attack by an adversary?}

To solve the above mentioned problem, we propose a frequency-hopping based relay-assisted key generation protocol with the following features: (i) the key generation phase between the nodes and the relay are executed on a given frequency band, which is public, and (ii) the broadcast phase is executed on a random frequency band from a dictionary of bands such that the adversary knows the dictionary but not the chosen band. 
The unique feature of the protocol is that a part of the secret key generated between Node-R and Node-A is used to pick the frequency band for the broadcast phase. While this idea facilitates both Node-R and Node-A to synchronously hop across the bands to recover the key, it also hides the chosen frequency band from the adversary, thereby evading the jamming attack with some probability. On the other hand, this idea results in loss of key-rate as some bits are used for frequency hopping. Keeping this strategy as the base model, we use state-of-the-art key generation protocols to formulate interesting optimization problems on how to maximize the key-rate of the protocol subject to a given reliability of recovery due to impairments from channel noise and jamming. Important resources such as power allocation for the probing and distribution phase, as well as the number of bits accounted for frequency hopping have been used as parameters for optimization. Finally, low-complexity solutions have been presented to obtain near-optimal solutions on the underlying parameters. We highlight that this is the first work of its kind in this space to address self-sustaining key generation under DoS threats in relay-assisted key generation models.

\begin{table}[h!]
  \begin{center}
    \caption{List of symbols}
    \label{tab:table1}
    \begin{tabular}{c|c} 
      \textbf{Notations} & \textbf{Definitions} \\
      \hline
      $\alpha$ & \text{bit allocation parameter} \\
      $\beta$ & \text{power allocation parameter} \\
      $\delta$ & \text{Mismatch rate} \\
      $c_{XR}$ & \text{LOS parameter for channel between Node-X and Node-R}\\
      $L$ & \text{Blocklength}\\
      $\epsilon$ & \text{Upper bound on error probability}
    \end{tabular}
  \end{center}
\end{table}

\section{Self-Sustainable Key Generation Model}

Consider a network, as shown in Fig. \ref{fig:network model}, where two nodes, say Node-A and Node-B, desire to establish a symmetric key by leveraging inherent randomness in their wireless channel. However, we assume that the channel between Node-A and Node-B is subject to shadowing or is hampered by a strong line-of-sight (LOS) component, thereby forcing them to seek assistance from a trusted node, say Node-R. The relay-assisted key generation protocol under consideration is divided into two phases, namely: the probing phase and the key distribution phase. Given that the three nodes do not have any prior credentials, we assume that the probing phase takes place on a pre-decided frequency band, which is public. Assuming a total power-budget of $P$ units, the probing and the broadcast phase use $\beta P$ and $(1-\beta)P$ fractions, respectively, where $0< \beta < 1$, is an underlying parameter for optimization. In the rest of the section, we explain the probing and distribution phases. 

\subsection{Probing Phase}

Each node transmits a probing symbol $\sqrt{\frac{\beta P}{3}}$ on the pre-decided frequency band, wherein the wireless channels between each pair is assumed to be quasi-static for $L$ channel-uses. As a result, the baseband symbols received at Node-A, Node-B, and Node-R on the $l$-th coherence-block are respectively given by,
\begin{eqnarray} 
y_{XZ}(l) = h_{XZ}(l) \sqrt{\frac{\beta P}{3}} + n_{XZ}(l),
\end{eqnarray}
where $X, Z \in \{A,B,R\}$ such that $X \neq Z$. For instance, with $X = A, Z = R$, $y_{AR}(l)$ is the received symbol at Node-R when Node-A transmits the probing symbol, $h_{AR}(l)$ is the channel from Node-A to Node-R, and $n_{AR}(l)$ is the additive white Gaussian noise (AWGN), distributed as $\mathcal{CN}(0, \gamma)$ at Node-R when Node-A sends the probing symbol. In this channel model, we assume that the complex baseband channel $h_{XR}$, for $X \in \{A, B\}$, is of the form $$h_{XR}(l) = \sqrt{c_{XR}}\left(\frac{1 + i}{2}\right) + \sqrt{1 - c_{XR}}g_{XR}(l),$$
wherein $\sqrt{c_{XR}}\left(\frac{1 + i}{2}\right)$, with $i = \sqrt{-1}$ captures the LOS component with $c_{XR} \in (0, 1)$, and $\sqrt{1 - c_{XR}}g_{XR}(l)$ captures the non-LOS (NLOS) component such that $g_{XR}(l) \sim \mathcal{CN}(0,1)$. We assume that the channels from Node-A to Node-R, and Node-R to Node-A are reciprocal, i.e., $h_{AR}(l) = h_{RA}(l)$, and similarly, between Node-B and Node-R. 

With the received symbols, each pair of nodes apply the state-of-the-art consensus algorithm in \cite{practical_key_generation}, wherein their inputs are unfolded real values corresponding to the complex received symbols over the $L$ coherence-blocks. In particular, as part of the consensus algorithm, each node feeds their corresponding sample to an quantizer with reference levels, $q_m$ and $q_p$ such that $q_{p} > q_{m}$. The reference levels are chosen such that a sample lying above $q_p$ will be treated as bit-1, a sample lying below $q_m$ is treated as bit-0, whereas a sample in between $q_m$ and $q_p$ is discarded. As a result, the samples on those index values that are not discarded by both the nodes are used to form a binary key. Based on the joint distribution of the real samples at the two participating nodes, $q_m$ and $q_p$ are chosen such that the mismatch rate between the keys at the nodes is bounded by a small number, say $\delta > 0$. Given two real samples, $x_{X}$ and $x_{R}$ at Node-X and Node-R, respectively, consensus probability, denoted by $p_\delta^{(XR)}$, refers to the probability that both samples are not discarded, thereby contributing to the formation of secret-key. Formally, $p_\delta^{(XR)}$ is given by  
	\begin{equation}
		\label{eq:consensus_prob}
		p_\delta^{(XR)} = \iint\limits_D f(x_X,x_R)\,dx_X\,dx_R,
	\end{equation}
such that $x_X$ and $x_R$ are the random variables that represent the real samples at Node-X and Node-R, respectively, and $f(x_X,x_R)$ is the joint density function between the two variables, and $D$ is the region of integration given by $D = \{(x_X, x_R): x_X \notin (q_m, q_p)\quad \& \quad x_R \notin (q_m, q_p)\}$.

After applying the consensus algorithm on the samples over $L$ coherence-blocks, Node-A and Node-R generate a symmetric key, $k_{AR}$ of length, $N_{AR}$, where the key length is a Binomial random variable whose average is $2Lp_\delta^{(AR)}$. Similarly, the other pair of nodes, Node-B and Node-R, also generate a symmetric key, denoted by $k_{BR}$, of length $N_{BR}$. Denoting the consensus probability for this pair of nodes by $p_\delta^{(BR)}$, its average key length is given as $2Lp_\delta^{(BR)}$.

\subsection{Key Distribution Phase}
After the key generation process, Node-R can broadcast the XOR version of $k_{AR}$ and $k_{BR}$ on the pre-decided frequency band in the $(L+1)$-th coherence block. Then Node-A can extract $k_{BR}$ from the broadcast message as it has the knowledge of $k_{AR}$, and finally, Node-A and Node-B can use $k_{BR}$ for securing their communication. However, in the scenario of a jamming attack on the broadcast phase, it is clear that Node-A will not be able to recover $k_{BR}$ if node-R continues to broadcast on the pre-decided band. To circumvent this problem, we adopt a frequency hopping technique to evade the jamming attack. To assist frequency hopping, we assume that all the nodes have the knowledge of a dictionary of frequency bands, denoted by $\mathcal{F}$, out of which one of them is chosen at random. Then Node-A and Node-R agree to carve out a few bits from the secret key $k_{AR}$. Based on these bits, a frequency band from the dictionary $\mathcal{F}$ is chosen in a deterministic manner and then used for the broadcast phase. Given that the jammer does not have the knowledge of $k_{AR}$, the specific choice of the frequency band will not be known, and thus the jamming attack is evaded. More importantly, Node-R can use the recovered bits from $k_{BR}$ as secret key with Node-B. Formally, for $0 < \alpha < 1$, which is a design parameter for optimization, let Node-A and Node-R, carve out $\floor{\alpha N_{AR}}$ bits for choosing the frequency band. With that the rest of the bits of $N_{AR}$, i.e. $\floor{(1 - \alpha)N_{AR}}$ will be used to confidentially share a part of $k_{BR}$ with Node-A. Using $k_{AR}$ and $k_{BR}$, Node-R obtains a new sequence $k_{XOR} \in \{0,1\}^{N_{XOR}}$, where $N_{XOR} = \mbox{min}\{\floor{(1 - \alpha)N_{AR}}, N_{BR}\}$, defined as
\begin{equation}
	\label{eq:XOR_operation}
	k_{XOR} = \left\{ \begin{array}{cccccccccc}
		\bar{k}_{AR} \oplus k_{BR}, & \mbox{ if } \floor{(1 - \alpha)N_{AR}} = N_{BR};\\
		\bar{k}_{AR} \oplus \bar{k}_{BR}, & \mbox{ if } \floor{(1 - \alpha)N_{AR}} < N_{BR};\\
		\bar{\bar{k}}_{AR} \oplus k_{BR}, & \mbox{ otherwise } \\
	\end{array}
	\right.
\end{equation}
where $\bar{k}_{AR}$ constitutes the first $\floor{(1 - \alpha)N_{AR}}$ components of $k_{AR}$, $\bar{k}_{BR}$ constitutes the first $\floor{(1 - \alpha)N_{AR}}$ components of $k_{BR}$, and $\bar{\bar{k}}_{AR}$ constitutes the first $N_{BR}$ components of $k_{AR}$.
Subsequently, $k_{XOR}$ is mapped to an $L$-length codeword $\mathbf{c} \in \mathcal{S} \subset \mathbb{C}^{L}$, and then broadcast to Node-A in the $(L+1)-$th coherence block on the new frequency band. Here, $\mathcal{S}$ denotes the chosen channel code of block-length $L$.\\ 
\indent On the new frequency band, Node-A receives
\begin{equation}
	\label{eq:receive_signal}
	y_{A}(n) = \sqrt{(1-\beta) P} h_{AR}(n) \mathbf{c}_{n} + n_{A}(n),
\end{equation}
for $1 \leq n \leq L$, where $\mathbf{c}_{n}$ is the $n$-th component of the codeword $\mathbf{c}$, $\mathbb{E}[|\mathbf{c}_{n}|^{2}] = 1$, and $n_{A}(n)$ 
represent the AWGN, distributed as $\mathcal{CN}(0, \gamma)$. Then, Node-A decodes to $\hat{\mathbf{c}} \in \mathcal{S}$ using an appropriate decoder, and then recovers $\hat{k}_{XOR}$ before extracting the shared secret-key $k_{BR}$ or $\bar{k}_{BR}$ as 
\begin{equation}
	\label{eq:XOR_decode}
	\hat{k}_{BR} = \left\{ \begin{array}{cccccccccc}
		\hat{k}_{XOR} \oplus \bar{k}_{AR}, & \mbox{ if } \floor{(1 - \alpha)N_{AR}} = N_{XOR};\\
		\hat{k}_{XOR} \oplus \bar{\bar{k}}_{AR}, & \mbox{ otherwise. }
	\end{array}
	\right.
\end{equation}

After Node-A extracts the key successfully, Node-A and Node-B reach to an agreement of using the secret key, $k_{BR}$ or $\bar{k}_{BR}$, which Node-B has obtained from the key generation phase. Thus, the above protocol is able to evade the jamming attack on the broadcast phase. In the next section, we discuss some important metrics that need to be optimized over the underlying parameters $\alpha, \beta$.

%

\section{Parameter Optimization}

It is desirable that the number of secret bits extracted between Node-A and Node-B is as large as possible. From the protocol, this can be achieved by allocating more power in the probing phase, i.e., with high $\beta$, as it increases the number of bits generated in a pair-wise manner at the relay. However, this reduces the power for key distribution phase thereby increasing the likelihood of unsuccessful delivery of key due to channel impairments. Another way to extract more secret bits from the protocol is to carve out fewer bits to select the frequency band for frequency hopping at Node-R, i.e., with small $\alpha$. However, fewer bits for generating the new frequency band increases the probability with which the jammer can chase the new frequency band. Thus, while
increasing $\beta$ and reducing $\alpha$ have positive effects on the key-rate, it may hamper the reliability with which the secret key is recovered at Node-A. Identifying this behaviour, we propose an optimization problem as a function of $\beta$ and $\alpha$ in the rest of this section. 

Suppose that the long-term statistics of the channel between Node-A and Node-R, and the channel between Node-B and Node-R, are identical. In such a case, for a given $\beta$ and $\alpha$, it is clear from  \eqref{eq:XOR_operation} that Node-R has an average of $(1 - \alpha)2p_\delta^{(AR)} L$ secret bits from the first $L$ coherence-blocks. Since these $(1 - \alpha)2p_\delta^{(AR)} L$ secret bits of $k_{BR}$ are communicated to Node-A in the broadcast phase, the average key-rate of the protocol is $(1 - \alpha)2p_\delta^{(AR)} \frac{L}{L+1}$ bits per coherence-block. Based on key distribution phase, the message communicated by Node-R may not reach Node-A when the instantaneous mutual information of the wireless link from Node-R to Node-A is less than the rate $\mathcal{R} = (1 - \alpha) 2p_\delta^{(AR)}$ bits per cu. We formally define the probability of such an event

\begin{IEEEeqnarray}{rcl}
	\label{eq:outage_expr_Marcum_Q}
	P^{(RA)}(O_C) & = & \mbox{Prob}\left(\mathcal{R} > \mbox{log}_{2}\left(1 + \frac{|h_{AR}|^2 (1-\beta)P}{\gamma}\right)\right) \nn\\
	& = & 1 - Q_1\left(\sqrt{\frac{2c_{AR}}{1 - c_{AR}}}, \sqrt{\frac{2(2^\mathcal{R} - 1)}{(1 - \beta)\rho(1 - c_{AR})}}\right)\nn\\
\end{IEEEeqnarray}
where signal-to-noise ratio (SNR), $\rho = \frac{P}{\gamma}$, and $Q_{1}(.,.)$ is a first-order Marcum-Q function, which is a complementary cumulative distribution function for the non-central Chi-squared random variable, $|h_{AR}|^2$. Note that the probability, $P^{(RA)}(O_C)$ given in the equation \eqref{eq:outage_expr_Marcum_Q} is a saddle point approximation of finite blocklength average error rate \cite{non_asymptotic_average_error_probability}, and this approximation is tight when the blocklength is of the order of few hundreds \cite{finite_blocklength}. Also, even when the instantaneous mutual information of the wireless link from Node-R to Node-A is more than the rate $\mathcal{R}$, the newly chosen frequency band may witness jamming energy when the attacker uniformly chooses one of the frequency bands in $\mathcal{F}$. The probability of such an event is given by
\begin{equation}
	\label{eq:probability_jamming}
	P^{(RA)}(O_J) = \frac{1}{2^{\floor{\alpha 2p_\delta^{(AR)}L}}}.
\end{equation}
Considering the above two events, we define effective average error probability, $P^{(RA)}(O)$, as the probability that the key is not successfully recovered by Node-A, given as 
\begin{equation}
	\label{eq:effective_probability}
	P^{(RA)}(O) = P^{(RA)}(O_C) + \left(1 - P^{(RA)}(O_C)\right)P^{(RA)}(O_J). 
\end{equation}
Taking the above discussions into account, we formulate a constrained optimization problem of maximizing the key-rate of the protocol subject to an upper bound on the effective average error probability, given by $P^{(RA)}(O) \le \epsilon$, for some $\epsilon > 0$. The first one is in \eqref{eq:solve_keyrate_ub} when there is no bound on the number of frequency bands available to form $\mathcal{F}$. The second one is in \eqref{eq:solve_keyrate_b} when the number of frequency bands available to form $\mathcal{F}$ is upper bounded by $2^{\kappa}$, for some $\kappa > 1$.
\begin{itemize}
	\item Unbounded number of frequencies: $\kappa = \infty$
\begin{small}
	\begin{align}
		\label{eq:solve_keyrate_ub}
			&\text {(P1):}
			& \underset{\alpha, \beta}{\text{argmax}}
			& \quad (1 - \alpha) 2p_\delta^{(AR)} \nn \\
			& \mbox{ such that } & &  P^{(RA)}(O) \leq \epsilon.
	\end{align}
 \end{small}
 \end{itemize}
  \begin{itemize}
	\item Bounded number of frequencies: $\kappa < \infty$
	\begin{equation}
		\label{eq:solve_keyrate_b}
		\begin{aligned}
			&\text {(P2):}
			& \underset{\alpha, \beta}{\text{argmax}}
			& \quad (1 - \alpha) 2p_\delta^{(AR)} \\
			& \mbox{ such that } & &  P^{(RA)}(O) \leq \epsilon; \\
			& & &  P^{(RA)}(O_J) \geq \frac{1}{2^\kappa}. \\
		\end{aligned}
	\end{equation}
\end{itemize}
Since $L$ is fixed, we have omitted the ratio $\frac{L}{L+1}$ in the objective functions of \eqref{eq:solve_keyrate_ub} and \eqref{eq:solve_keyrate_b}. In the following section,
we provide a solution for the optimization problems, by first
deriving a mathematical expression for the objective function

\subsection{Challenges in Solving (P1) and (P2)}

Towards solving the optimization problems in (P1) and (P2), it is first important to express the objective functions
and the constraints as a function of the underlying parameters
$\beta$ and $\alpha$. Along that direction, we identify that the consensus probability $p_\delta^{(AR)}$, which is a function of $\beta$, appears in both the objective functions and the constraints. However, from its definition in \eqref{eq:consensus_prob}, expressing $p_\delta^{(AR)}$ as a function of $\beta$ needs us to compute an area of a bi-variate Gaussian distribution over rectangular regions, which is not known in closed-form hitherto \cite{kotz}. Although a form of Taylor series approximation can be applied to \eqref{eq:consensus_prob} after some algebraic manipulations, the approximation will only be a function of $q_{p}$ and $q_m$. Since the variables $q_{p}$ and $q_m$ are also functions of $\beta$, and there is no tractable expression to relate them, this Taylor series approximation \cite{olson} can only be obtained through empirical results. These intractable problems throw challenges towards elegantly solving the problem statements in (P1) and (P2). To circumvent this problem, in the next section, we present a regression based strategy to analytically solve (P1) and (P2), and gain insights into the optimal choice of $\beta$ and $\alpha$.  

\begin{figure}[ht!]
	\begin{center}
        \includegraphics[width=\linewidth]{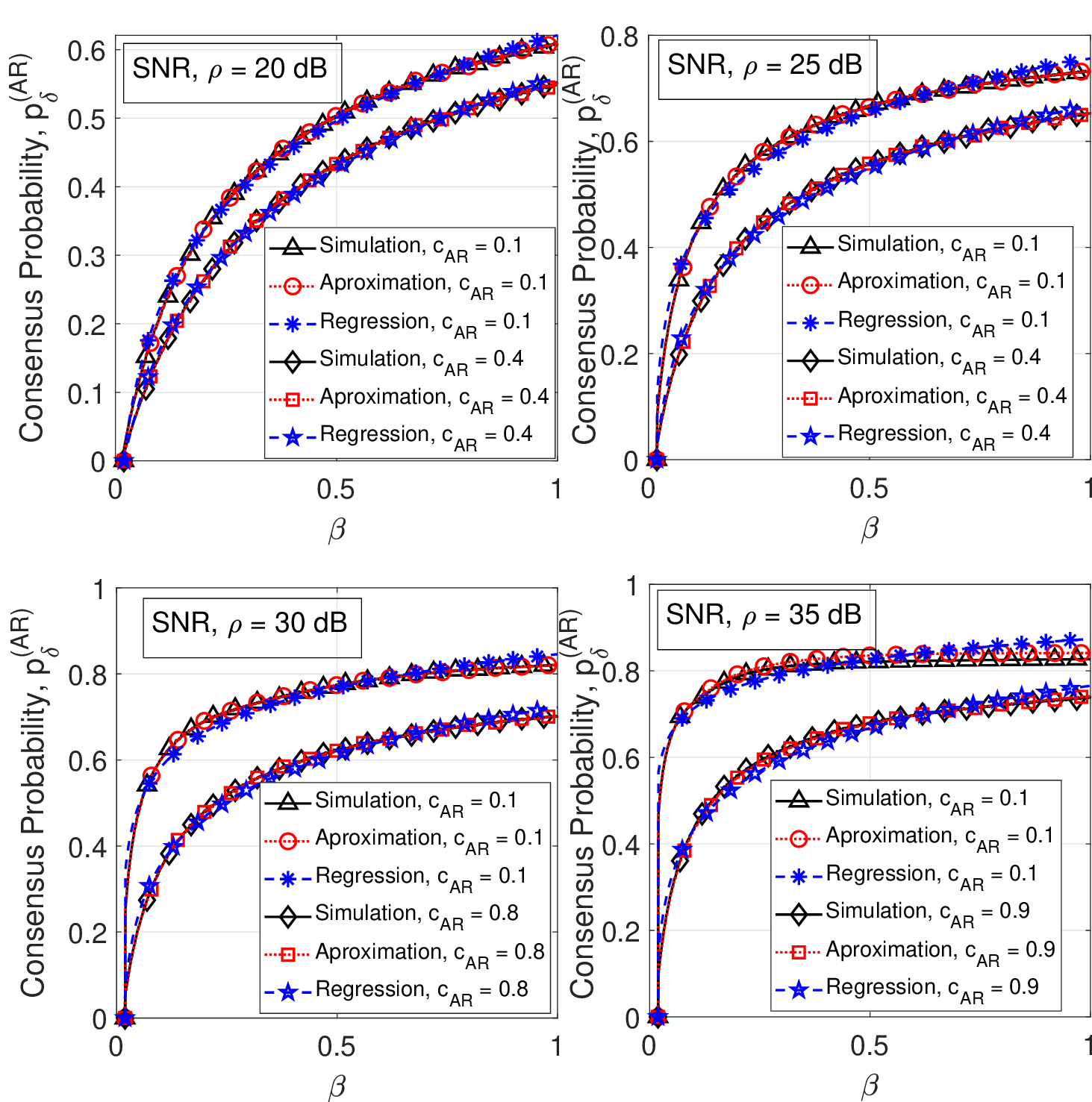}
	\end{center}
	\vspace{-0.3cm}
	\caption{Comparison of simulation of consensus probability with its approximation and non-linearly regressed expression for different values of SNR, $\rho$ and LOS parameter, $c_{AR}$}.
    \label{fig:consensus_probability_curve_fitting}
\end{figure}


\section{Regression Based Key-Rate Analysis}

For a given SNR, LOS component and the mismatch rate, we observe from a series of experiments that $p^{(AR)}_\delta$ behaves in a logarithmic fashion as a function of $\beta$. With this insight, we fit a mathematically tractable model by applying a non-linear regression algorithm. In particular, we first express $p^{(AR)}_\delta$ as
\begin{equation}
	\label{eq:consensus_regression}
	p^{(AR)}_\delta \approx \tilde{p}_\delta^{(AR)} \triangleq p_1 + p_2\mbox{log}_2(\beta + p_3),
\end{equation}
where, $p_j \in \mathbb{R}_+, j = 1,2$, and $p_3 \in \mathbb{R}$ are the variables which are regressed, $\beta$ is the independent variable, and $\tilde{p}_\delta^{(AR)}$ denotes the approximation on $p^{(AR)}_\delta$. For a given LOS component, SNR, and a mismatch rate, we empirically obtain the consensus probability values for some values of $\beta \in (0, 1)$ through simulations. Subsequently, we run a non-linear regression algorithm to estimate the values of $p_{1}, p_{2}$ and $p_{3}$. To validate the accuracy of this approach, in Fig. \ref{fig:consensus_probability_curve_fitting}, we present consensus probability as a function of $\beta$ for various values of SNR and LOS parameters. The ones corresponding to simulations refer to consensus probability obtained through Monte-Carlo simulations, whereas the ones corresponding approximation refer to the expressions obtained using Taylor series based approximations discussed in the previous section. The plots confirm that the regression-based expression is reasonably accurate to apply the same when solving the (P1) and (P2). 

While \eqref{eq:consensus_regression} can be used in place of $p^{(AR)}_\delta$ in the objective functions and the constraints of (P1) and (P2), we observe that the constraints space still contain first-order Marcum-Q functions, as we observe in \eqref{eq:outage_expr_Marcum_Q} and \eqref{eq:effective_probability}. Since tackling the first-order Marcum-Q function is analytically challenging, we use the following theorem to replace the Marcum-Q functions by their approximation in the constraints space of (P1) and (P2).  

\begin{theorem}
	At high SNR regime, power allocation parameter, $\beta \in (0,1)$, and LOS parameter, $c_{AR} \in (0,1)$, the first-order Marcum Q-function in \eqref{eq:outage_expr_Marcum_Q} can be approximated as
	\begin{equation}
		\label{eq:marcum_approx}
		Q_1(a,b) \approx \tilde{Q}_1(a,b) \triangleq 1 + e^{-\frac{a^2}{2}}\{e^{-\frac{b^2}{2}} - 1\},
	\end{equation}
where $a = \sqrt{\frac{2c_{AR}}{1 - c_{AR}}}$, and $b = \sqrt{\frac{2(2^\mathcal{R} - 1)}{(1 - \beta)\rho(1 - c_{AR})}}$
\end{theorem}

Using the approximation in \eqref{eq:marcum_approx}, we approximate \eqref{eq:outage_expr_Marcum_Q} as
\begin{equation}
	\label{eq:outage_expr_Marcum_Q_approx}
	\tilde{P}^{(RA)}(O_C) = 1 - \tilde{Q}_1(a,b) \triangleq e^{-\frac{a^2}{2}}\{1 - e^{-\frac{b^2}{2}}\}.
\end{equation}
Using \eqref{eq:outage_expr_Marcum_Q_approx} in \eqref{eq:effective_probability}, the effective average error probability is

\begin{small}
\begin{eqnarray}
	\label{eq:effective_probability_approx}
	\tilde{P}^{(RA)}(O) & = & \tilde{P}^{(RA)}(O_C) + \left(1 - \tilde{P}^{(RA)}(O_C)\right)\tilde{P}^{(RA)}(O_J) \nonumber\\
 & < & \tilde{P}^{(RA)}(O_C) + \tilde{P}^{(RA)}(O_J),
\end{eqnarray}
\end{small}

\noindent where $\tilde{P}^{(RA)}(O_C)$, $\tilde{P}^{(RA)}(O_J)$, and $\tilde{P}^{(RA)}(O)$ denote the approximations on $P^{(RA)}(O_C)$, $P^{(RA)}(O_J)$, and $P^{(RA)}(O)$  respectively. Using the approximations on the consensus probability in \eqref{eq:consensus_regression}, and effective average error probability in \eqref{eq:effective_probability_approx}, we reformulate the optimization problems in (P1) and (P2) as (P3) and (P4), respectively shown below:

\begin{itemize}
	\item Unbounded number of frequencies: $\kappa = \infty$
	\begin{align}
		\label{eq:p3}
			&\text {(P3):}
			& \underset{\alpha, \beta}{\text{argmax}}
			& \quad (1 - \alpha) 2\tilde{p}_\delta^{(AR)} \\
			& \mbox{ s. t. } & &  \tilde{P}^{(RA)}(O_C) + \tilde{P}^{(RA)}(O_J) \leq \epsilon.
		\end{align}
	\item Bounded number of frequencies: $\kappa < \infty$
	\begin{equation}
		\label{eq:p4}
		\begin{aligned}
			&\text {(P4):}
			& \underset{\alpha, \beta}{\text{argmax}}
			& \quad (1 - \alpha) 2\tilde{p}_\delta^{(AR)} \\
			& \mbox{ s. t. } & &  \tilde{P}^{(RA)}(O_C) + \tilde{P}^{(RA)}(O_J) \leq \epsilon \\
			& & &  \tilde{P}^{(RA)}(O_J) \geq \frac{1}{2^\kappa}. \\
		\end{aligned}
	\end{equation}
\end{itemize}

Unlike (P1) and (P2), it is clear that their counterparts (P3) and (P4) are more analytically tractable. In the following lemma, we present some interesting results on the objective functions and the constraints in (P3) and (P4).

\begin{figure}
\begin{center}
		\includegraphics[scale = 0.4]{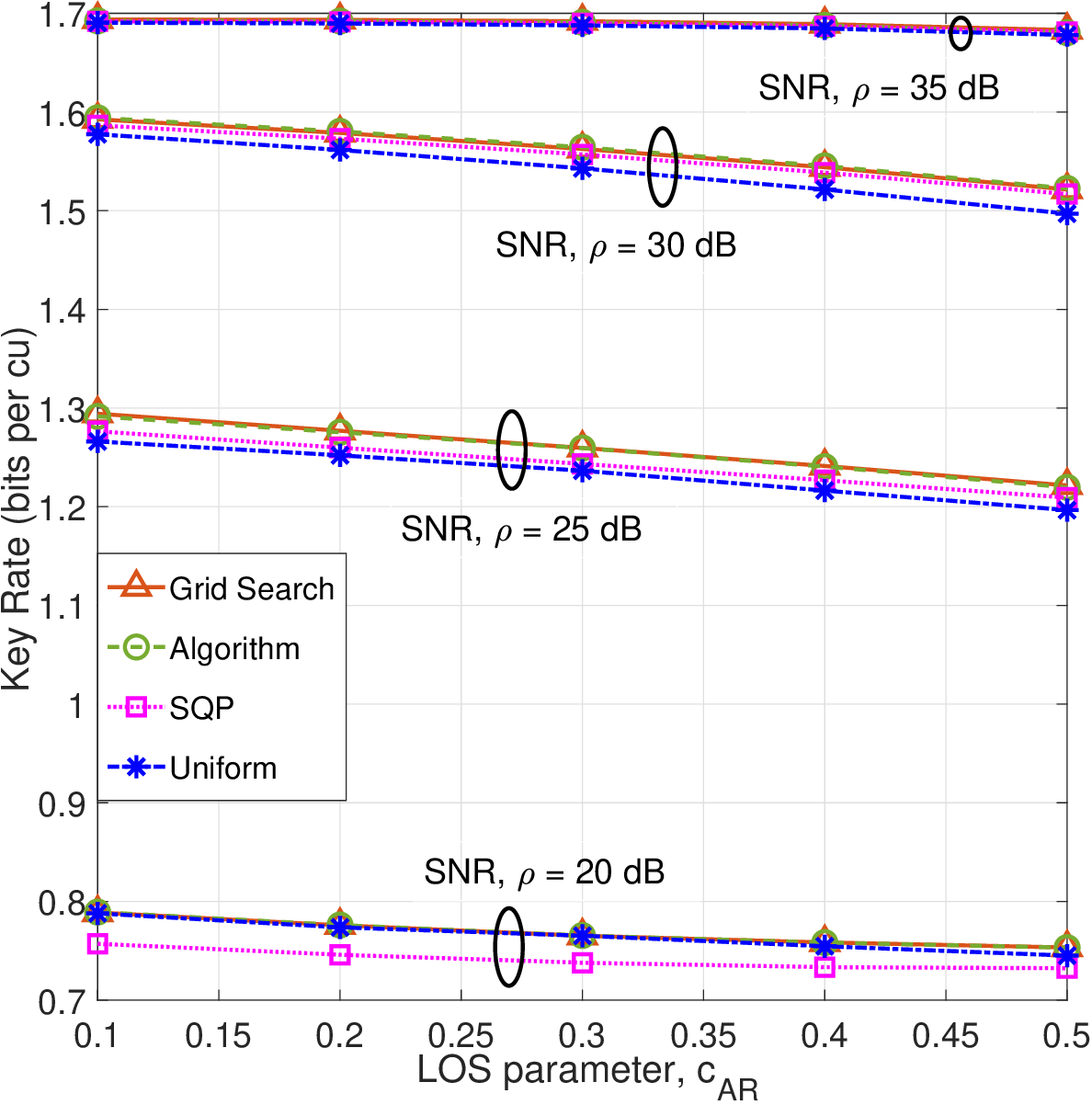}
  \end{center}
		\caption{Key-rates offered by the proposed self-sustainable key-generation protocol when $\alpha$ and $\beta$ are optimized using various methods: Grid search is the exhaustive search over $\alpha$ and $\beta$, SQP is the sequential quadratic programming applied on (P3) and (P4), Uniform is a variant of our algorithm wherein the initial value of $\beta$ is chosen from $(0, 1)$ with uniform distribution.}
		\label{Fig: key_rate_ub}
  \end{figure}

 \begin{figure}
 \begin{center}
		\includegraphics[scale = 0.4]{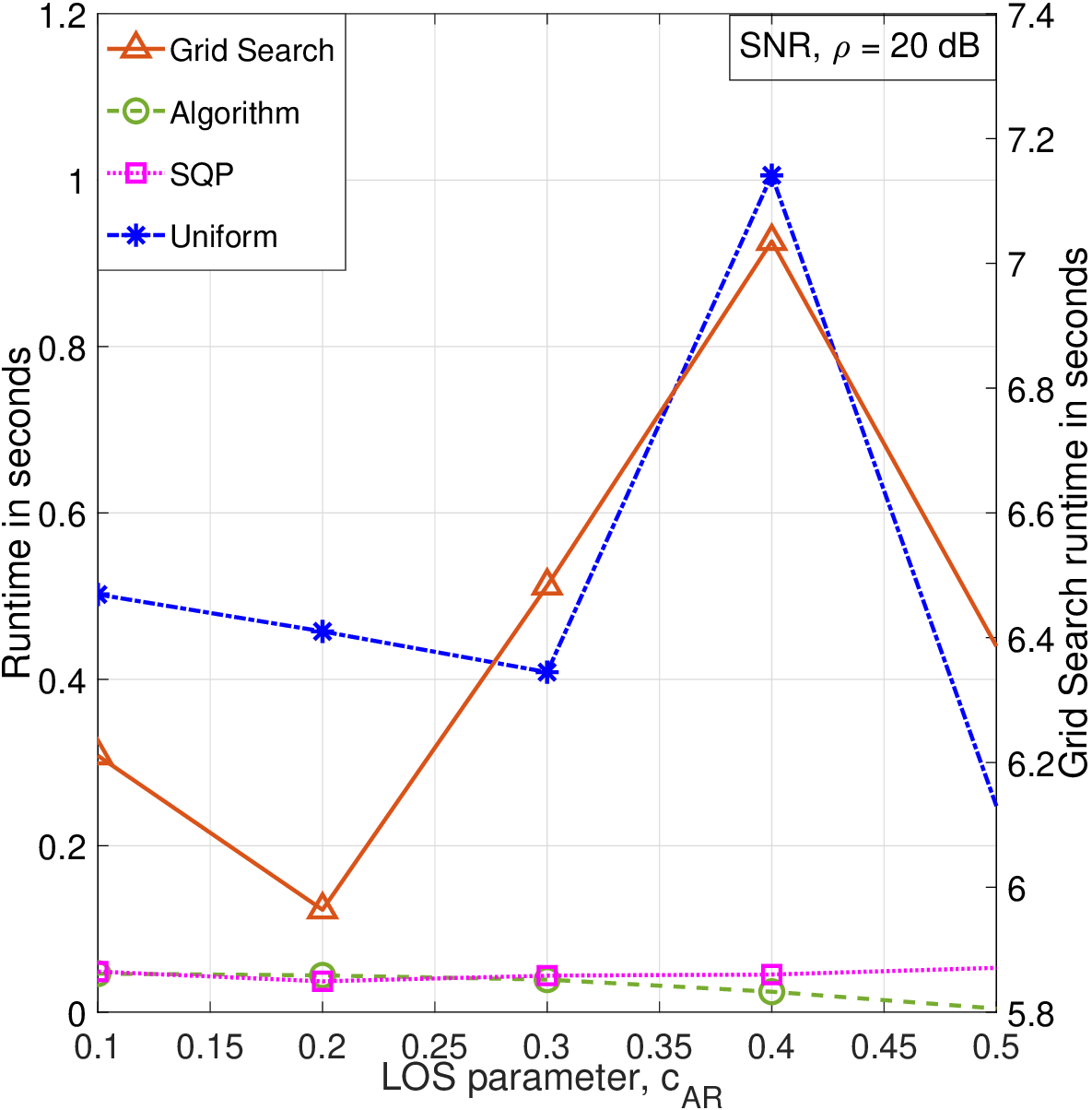}
  \end{center}
		\caption{Run-times of various algorithms used to solve (P3) and (P4). The run-times of grid search specifically plotted on the right-hand-side of y-axis due to significant order difference with respect to the other modules.}
		\label{Fig: rt_ub}
\end{figure}

\begin{figure}
\begin{center}
\includegraphics[scale = 0.385]{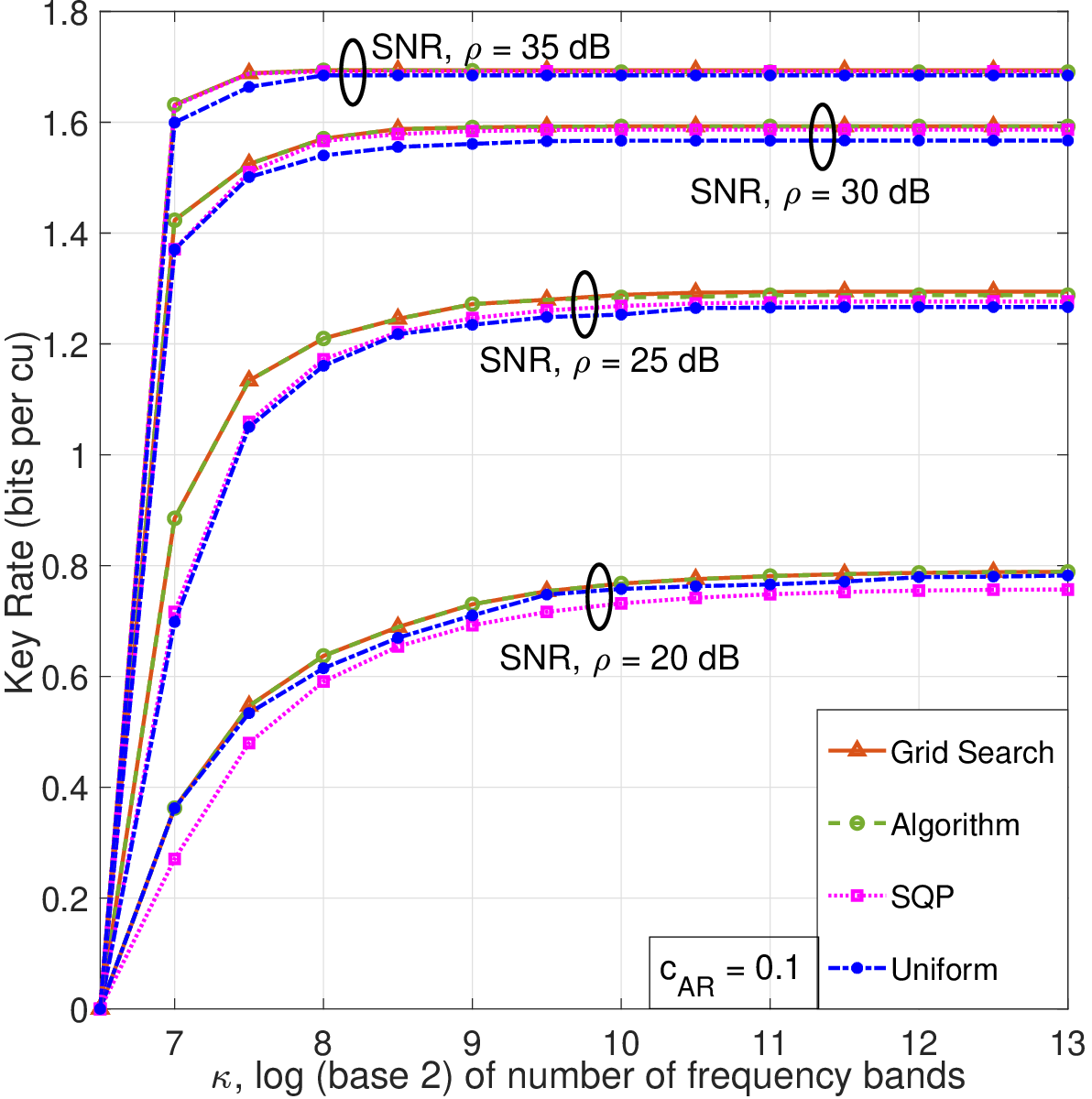}
  \end{center}
		\caption{Comparison of key-rates when $\alpha$ and $\beta$ are optimized using various methods to solve (P4). These plots are when $\kappa$ is bounded.}
		\label{Fig: key_rate_b}
  \end{figure}
\begin{figure}
\begin{center}
		\includegraphics[scale = 0.4]
  {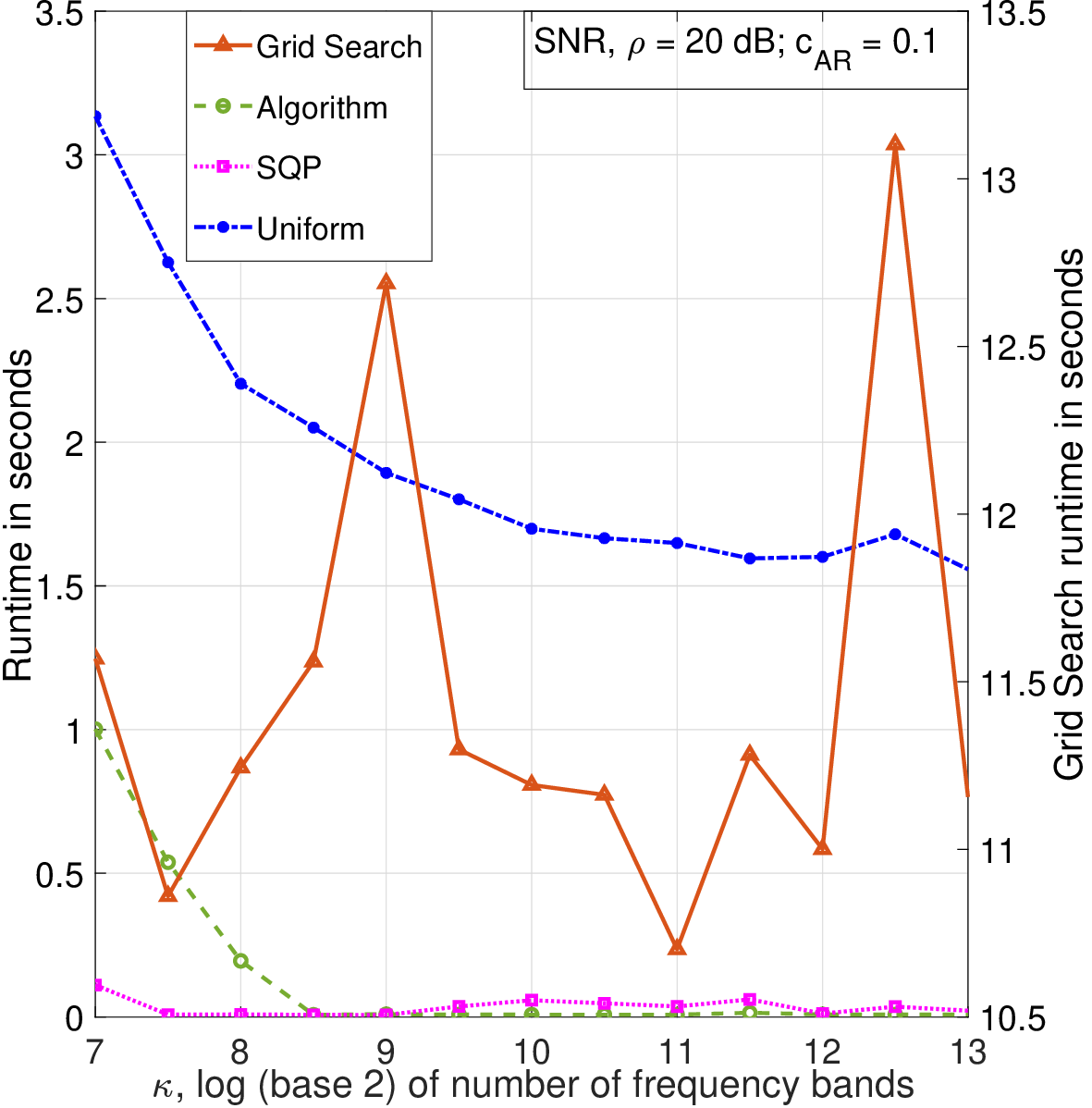}
  \end{center}
		\caption{Run-times of various algorithms used to solve (P4), i.e., when $\kappa$ is bounded.}
		\label{Fig: rt_b}
\end{figure}

\begin{lemma}
\label{lemma: characertistics_objective_func}
For a given SNR, LOS component, and a mismatch rate, the following statements can be proved. 
\begin{itemize}
\item (R1) For a given $\alpha$, the key-rate $(1 - \alpha) 2\tilde{p}_\delta^{(AR)}$ is an increasing function of $\beta$
\item (R2) For a given $\beta$, the key-rate $(1 - \alpha) 2\tilde{p}_\delta^{(AR)}$ is a decreasing function of $\alpha$
\item (R3) For a given $\beta$, $\tilde{P}^{(RA)}(O_C)$ and $\tilde{P}^{(RA)}(O_J)$ are decreasing functions of $\alpha$.
\item (R4) In the absence of DoS attack, we have $\alpha = 0$, and the optimal value of $\beta$ to solve (P3) and (P4) can be obtained in closed-form. 
\end{itemize}
\end{lemma}

Using the above analytical results, we present a low-complexity algorithm to solve (P3) an (P4) in the presence of a DoS attack on the broadcast phase.  

\subsection{Low-Complexity Algorithm to Solve (P3) and (P4)}

We present a low-complexity algorithm for solving the optimization problem in (P3). A variant of the same can be applied to solve (P4) with an additional constraint on the probability of jamming. First, the set of feasible pairs of $(\beta, \alpha)$ that satisfy the constraint space must be identified. However, a feasible solution does not always exist for any upper bound on the error probability, $\epsilon$. Therefore, the rest of the discussion is applicable only when a feasible solution exists. From (R3) of Lemma~\ref{lemma: characertistics_objective_func}, it is clear that for a given $\beta$, the minimum value of $\alpha$ in the feasible region maximizes the objective function. Also, this value of $\alpha$ can be computed using the constraint function. Therefore, given that $\alpha$ is function of $\beta$, the main objective in designing a low-complexity algorithm is to determine an appropriate value of $\beta$. In the proposed idea, which is as given in Algorithm 1, we start with the initial value of $\beta_{init}$ which maximizes (P3) for $\alpha = 0$. Subsequently, we compute the minimum value $\alpha$ that satisfies the constraint, and then note down the corresponding key-rate value. Then, we increment the value of $\beta$ from $\beta_{init}$ with a given step-size, compute its corresponding $\alpha$, and then note down its key-rate. We continue this process until a local-maxima is reached. That would be the output of Algorithm 1. Similarly, the same procedure is repeated by decreasing from $\beta_{init}$ until a local maxima on key-rate is reached. Note that a variant of Algorithm 1 can be used to represent this direction, with the only difference that in line 6, we must decrement $\beta$. Finally, the maximum of the two local-maxima are picked as the output of our method. 

\begin{algorithm}
\caption{Outage-driven greedy search}
\label{alg:practical_algorithm_unbounded_1}
\begin{algorithmic}[1]
\Require $L$, $\rho$, $c_{AR}$, $p_1$, $p_2$, and $p_3$
\Ensure Optimal values of the objective function and the underlying solutions $\alpha$ and $\beta$
\State \textbf{Initialize:} Set $\alpha = 0$ $\mathcal{R}_{old} = 0$, $\tilde{P}^{(RA)}(O_J) = 1$
\State Compute $\beta_{init}$ using (R4) of Lemma~\ref{lemma: characertistics_objective_func} and then assign $\beta = \beta_{init}.$
\State Using $\beta$, determine the minimum value of $\alpha$, that satisfies the constraint $\tilde{P}^{(RA)}(O_C) + \tilde{P}^{(RA)}(O_J) \le \epsilon$ \label{find_alpha} using (R3) of Lemma~\ref{lemma: characertistics_objective_func}
\State Compute the key-rate, and save it in $R_{new}$. Also store the corresponding $\alpha$ and $\beta$ values.
        \If{$\mathcal{R}_{new} > \mathcal{R}_{old}$}
            \State $\mathcal{R}_{old} = \mathcal{R}_{new}$, increment $\beta$, and goto Step: \ref{find_alpha} \label{step_beta}
        \Else
            \State \textbf{break}
        \EndIf
    \State Save $\mathcal{R}_{old}$ as the the local-maxima of key-rate, and also store the corresponding values of $\alpha$, and $\beta$.
\end{algorithmic}
\end{algorithm}

\begin{figure}[ht!]
	\begin{center}
        \includegraphics[scale = 0.5]{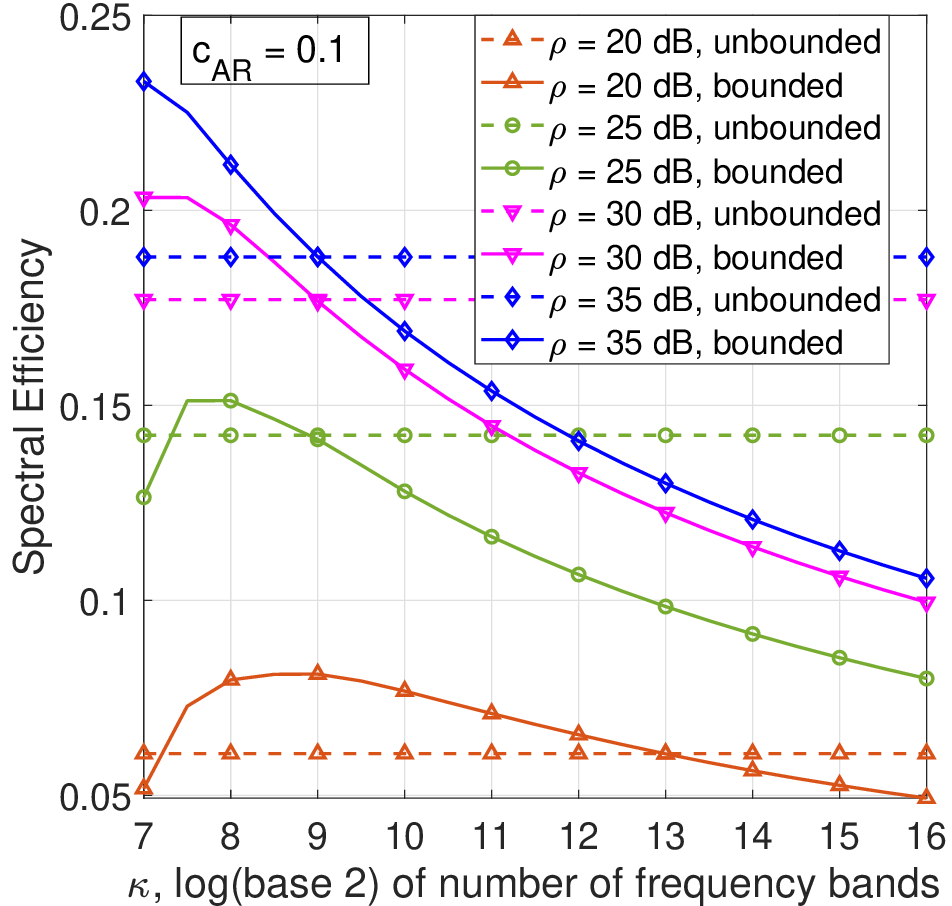}
	\end{center}
	\vspace{-0.1cm}
	\caption{Comparing the spectral-efficiency in key-rate per number of bits used for frequency-hopping when solving (P3) and (P4). Plots suggests that solving (P4) with an appropriate value of $\kappa$ yields maximum spectral-efficiency.}
    \label{fig:spectral_efficiency}
\end{figure}

\subsection{Simulation Results}

In this section, we present simulation results that demonstrate the efficacy of the proposed algorithm in comparison with the exhaustive or grid search, sequential quadratic programming (SQP), and a variant of the proposed algorithm with `uniformly' choosing initial value. Here, the initial value of $\beta$ is chosen with uniform distribution in the interval $(0, 1)$. We use $L = 200$ to generate the simulation results. 
In Fig. \ref{Fig: key_rate_ub}, we plot the optimal key-rates offered by the above four techniques for different values of $c_{AR}$, and SNR, $\rho$ when solving the optimization problem (P3) for $\epsilon = 10^{-2}$, and a mismatch rate of $\delta = 10^{-3}$. The plots confirm that the proposed algorithm provides near-optimal solutions. We emphasize there is no constraint on the number of frequency bands used for frequency-hopping when solving (P3). As a result the size of $\mathcal{F}$ is determined as $2^{\lfloor 2(\alpha)L\tilde{p}_\delta^{(AR)}\rfloor}$. For the parameters used to generate Fig. \ref{Fig: key_rate_ub}, we also present the average run-times of the four methods in Fig. \ref{Fig: rt_ub}. The plots show that the proposed algorithm is computationally more efficient than the others.

For solving (P4), we run the simulations with the additional constraint on the number of frequency bands available for frequency-hopping. From Fig. \ref{Fig: key_rate_b}, we observe that the optimal key-rate increases by increasing the number of available frequencies and then saturates. This behaviour is attributed to the fact that the channel outage probability dominates the jamming probability after a certain value of $\kappa$. In Fig. \ref{Fig: rt_b}, we also plot the run-times of each technique, and from the plots, we observe that the run-time of the proposed algorithm is way too low in comparison with that of exhaustive search. However, when $\kappa$ is less than or equal to 8, the run-time of the proposed algorithm is slightly higher than that of SQP. This behavior is attributed to the fact that for $\kappa \le 8$, the maximum feasible value of $\beta$ is much smaller than the initial value of $\beta$. As a result, the search space of it is large, which contributes to higher run-times. However, at higher values of $\kappa$, we observe that the proposed algorithm has higher computational efficiency than SQP.

When solving (P4), it is to be noted that spectral-efficiency of the protocol decreases as $\kappa$ increases. As a result, for a fair comparison, we compare the key-rates of the protocol normalized by the number of bits used to select the frequency band for hopping. Such a metric is presented in Fig. \ref{fig:spectral_efficiency} as a function of $\kappa$. In the same plot, we also use a horizontal line to present the spectral-efficiency of the protocol when solved using (P3). The plots show that there exists an optimal value of $\kappa$ that maximizes the spectral-efficiency to achieve the same reliability of $\epsilon$ in the broadcast phase.

\section{Summary}

In this work, we have addressed a DoS attack on relay-assisted key generation, wherein two wireless nodes take the help of a trusted relay node to generate secret-keys. In particular, we have considered a model wherein the active adversary injects jamming energy on the distribution phase of the protocol. To mitigate this jamming attack, we have proposed a self-sustainable key generation model, wherein a portion of the secret bits generated in the key generation phase is employed to implement frequency-hopping in the broadcast phase in order to evade the jamming attack. Using this countermeasure, we have proposed several optimization problems on how to maximise the key-rate of the protocol so as to ensure reliable delivery of the secret-keys. Several low-complexity solutions on the optimization problems are also presented, along with extensive simulations to justify their efficacy. 

Given that frequency-hopping is used in the broadcast phase to evade the jamming attack, the total number of frequency bands that are available must be noted. Recall that frequency-hopping across all the available frequency bands may not maximise the spectral-efficiency. As a result, the proposed optimization problem in (P4) must be solved to obtain the corresponding values of $\beta$ and $\alpha$ for various values of $\kappa$ upto the total number of available frequency bands. Finally, the value of $\kappa$ that provides the highest spectral-efficiency in key-rate must be used in practice.


\end{document}